\documentclass[draft,jgrga]{agutex}
\usepackage{graphicx}
\setkeys{Gin}{draft=false}
\authorrunninghead{REIMOND AND BAUR}

\titlerunninghead{GRAVITATIONAL POTENTIAL OF SMALL BODIES}

\authoraddr{Corresponding author: S. Reimond,
Space Research Institute, Austrian Academy of Sciences,
Schmiedlstra{\ss}e 6, 8042 Graz, Austria.
(stefan.reimond@oeaw.ac.at)}

\begin{document}

%% ------------------------------------------------------------------------ %%
%
%  TITLE
%
%% ------------------------------------------------------------------------ %%

\title{Spheroidal and ellipsoidal harmonic expansions of the gravitational potential of Small Solar System Bodies. Case study: Comet 67P/Churyumov-Gerasimenko}
%

%% ------------------------------------------------------------------------ %%
%
%  AUTHORS AND AFFILIATIONS
%
%% ------------------------------------------------------------------------ %%

%Use \author{\altaffilmark{}} and \altaffiltext{}

\authors{Stefan Reimond\altaffilmark{1},Oliver Baur\altaffilmark{1}\altaffilmark{,2}}

\altaffiltext{1}{Space Research Institute, Austrian Academy of Sciences, Graz, Austria.}
\altaffiltext{2}{Now at Airbus Defence and Space GmbH, Navigation and Apps Programmes, Ottobrunn, Germany.}

%% ------------------------------------------------------------------------ %%
%
%  ABSTRACT
%
%% ------------------------------------------------------------------------ %%

\begin{abstract}
Gravitational features are a fundamental source of information to learn more about the interior structure and composition of planets, moons, asteroids and comets. Gravitational field modeling typically approximates the target body with a sphere, leading to a representation in spherical harmonics. However, small celestial bodies are often irregular in shape, and hence poorly approximated by a sphere. A much better suited geometrical fit is achieved by a tri-axial ellipsoid. This is also mirrored in the fact that the associated harmonic expansion (ellipsoidal harmonics) shows a significantly better convergence behavior as opposed to spherical harmonics. Unfortunately, complex mathematics and numerical problems (arithmetic overflow) so far severely limited the applicability of ellipsoidal harmonics. In this paper, we present a method that allows expanding ellipsoidal harmonics to a considerably higher degree compared to existing techniques. We apply this novel approach to model the gravitational field of comet 67P, the final target of the Rosetta mission. The comparison of results based on the ellipsoidal parameterization with those based on the spheroidal and spherical approximations reveals that the latter is clearly inferior; the spheroidal solution, on the other hand, is virtually just as accurate as the ellipsoidal one. Finally, in order to generalize our findings, we assess the gravitational field modeling performance for some 400 small bodies in the solar system. From this investigation we generally conclude that the spheroidal representation is an attractive alternative to the complex ellipsoidal parameterization on the one hand, and the inadequate spherical representation on the other hand.
\end{abstract}

%% ------------------------------------------------------------------------ %%
%
%  BEGIN ARTICLE
%
%% ------------------------------------------------------------------------ %%

% The body of the article must start with a \begin{article} command
%
% \end{article} must follow the references section, before the figures
%  and tables.

\begin{article}

%% ------------------------------------------------------------------------ %%
%
%  TEXT
%
%% ------------------------------------------------------------------------ %%

\section{Introduction}
\label{sec:intro}
Small Solar System Bodies \citep{IAU2006} such as asteroids or comets produce in their surroundings heavily irregular gravitational fields. Both the nonsphericity and the massively roughened surfaces of these bodies present a challenge when developing accurate gravitational field models. With the increased number of dedicated space missions to extraterrestrial bodies the number of methodologies to face this particular challenge rose accordingly \citep{Scheeres2012}.

Direct modeling techniques rely on mass distribution assumptions in the interior of the body. The most innovative approach in this framework is the polyhedron method developed by \citet{Werner1994}. Apart from the density uncertainty, the accuracy of the analytically derived gravitational effects is exclusively limited to the quality and resolution of the three-dimensional shape model. One of the advantages of the polyhedron method is the possibility to study and predict flight dynamics of spacecraft in close proximity to the body. 

In contrast to that, inversion techniques make use of measurements taken outside the attracting object (e.g. spacecraft trajectory perturbations) to draw conclusions about its internal composition \citep{Seeber2003}. The gravitational potential can be expressed in terms of infinite harmonic series using a set of global basis functions. Convergence of these series is guaranteed outside a mass-enclosing reference surface, also referred to as Brillouin surface. Note that in the following, we will use these two terms interchangeably. 
When parameterized in spherical coordinates, the corresponding series is called spherical harmonics (SH) expansion and the geometrical reference surface is the Brillouin sphere \citep{Hobson2012}. SH are adequate for representing the gravitational field of planetary or other sphere-like bodies. When it comes to gravitational field modeling of irregularly shaped asteroids or comets, however, SH might become a sub-optimal choice due to the large discrepancy between the mass-enclosing reference sphere and the actual shape of the body. If the computation points are located near the surface of the body (e.g. during close encounters of space probes) and thus possibly inside the reference sphere, the harmonic series might diverge or converge to a value that does not represent the true value of the gravitational potential \citep{Hofmann-Wellenhof2006}. 

In order to mitigate this problem the use of alternative parametrizations has been suggested. Specifically, ellipsoidal harmonics (EH) have been investigated in great detail during the last two decades for this purpose \citep{Garmier2001,Garmier2002,Dechambre2002,Dassios2012,Park2014,Hu2015}. The underlying tri-axial reference ellipsoid (or Brillouin ellipsoid) approximates odd shapes decisively better than the sphere. Consequently, the region of convergence increases and close-range evaluations are possible. Unfortunately, the mathematical and numerical complexity involved in computing the basis functions of this harmonic expansion, i.e. the Lam\'{e} functions of the first and second kind, considerably reduces the usability of EH. One of the most serious limitations of EH are arithmetic over- and underflow errors caused by degreewise increasing orders of magnitude of the basis function values. These large numerical quantities impede the accurate computation of expansions higher than degree 10 to 15 \citep{Bardhan2012}.

In addition to SH and EH, Laplace's equation can also be solved using spheroidal coordinates \citep{Byerly2003}. A spheroid is a bi-axial ellipsoid which is obtained by rotating an ellipse about one of its semi-axes. Contracted bodies like the Earth are well approximated by an oblate spheroid. Therefore, oblate spheroidal harmonics (OH) have already been used extensively in geopotential modeling \citep{Thong1989,Sanso1993}. Prolate spheroidal harmonics (PH) are effective for modeling elongated celestial bodies \citep{Fukushima2014}. The spheroidal expansions combine the advantages of SH and EH, i.e. simple mathematics and a good geometric fit. The respective reference surfaces are denoted accordingly as the oblate and the prolate Brillouin spheroid.

The aim of this paper is threefold: First we present a method to increase the expansion degree of EH by making use of logarithmic expressions. We demonstrate that numerically stable results can be achieved up to at least degree 500. Second, we investigate the suitability of the SH, OH, PH and EH models for representing the gravitational field of comet 67P/Churyumov-Gerasimenko. We rely on simulation strategies to assess the performances of the various parametrizations. Third, we more generally assess the quality of the aforementioned parametrizations by a large-scale investigation including some 400 celestial bodies.

\section{Gravity field parametrizations}
For an arbitrary curvilinear and orthogonal reference system with coordinates $\xi_1,\xi_2,\xi_3$ Laplace's equation is expressed by \citep{Dassios2012}
\begin{eqnarray}
  \Delta V \left( \xi_1,\xi_2,\xi_3 \right) &= & \frac{1}{h_1 h_2 h_3} \left[ \frac{\partial}{\partial \xi_1} \left( \frac{h_2 h_3}{h_1} \frac{\partial V}{\partial \xi_1} \right) \right. \nonumber \\
  && + \frac{\partial}{\partial \xi_2} \left( \frac{h_1 h_3}{h_2} \frac{\partial V}{\partial \xi_2} \right) + \left. \frac{\partial}{\partial \xi_3} \left( \frac{h_1 h_2}{h_3} \frac{\partial V}{\partial \xi_3} \right) \right] \nonumber \\
  && = 0.
\label{eq:eq1}
\end{eqnarray}
Herein, $V$ is the gravitational potential at location $\xi_1,\xi_2,\xi_3$ and the scale factors $h_1,h_2,h_3$ are the roots of the metric coefficients. Separation of variables yields three ordinary differential equations:
\begin{equation}
V \left( \xi_1,\xi_2,\xi_3 \right) = \zeta \left( \xi_1 \right) \eta \left( \xi_2 \right) \theta \left( \xi_3 \right).
\label{eq:eq2}
\end{equation}
The solutions of Laplace's equation are harmonic functions. As for any linear homogeneous differential equation, $V$ can be written as a linear combination of individual solutions $V_n$ \citep{Haberman2013}:
\begin{equation}
V = \sum_{n=0}^{\infty} V_n, \quad V_n = \zeta_n \eta_n \theta_n.
\label{eq:eq3}
\end{equation}
For the sake of simplicity we dropped the dependencies on the coordinates in eq.~\ref{eq:eq3}. Table~\ref{tab:tab1} summarizes the exterior solutions of eq.~\ref{eq:eq3} in spherical, spheroidal and ellipsoidal coordinates. The following subsections give a brief exposition of their usage for gravitational field modeling.

\subsection{Spherical harmonics}
The spherical coordinate system comprises one radial component (the euclidean distance $r$) and two angular coordinates (in literature often introduced as colatitude $\vartheta$ and longitude $\lambda$). Expansion of the gravitational potential in spherical harmonics reads \citep{Torge2012}
\begin{eqnarray}
   V \left( r, \vartheta, \lambda \right) &= & \frac{GM}{R} \sum_{n=0}^{\infty} \sum_{m=0}^{n} \left( \frac{R}{r} \right)^{n+1} \overline{P}_{nm} \left( \cos \vartheta \right) \nonumber \\
   && \times \left[ \overline{c}_{nm} \cos m \lambda + \overline{s}_{nm} \sin m \lambda \right],
\label{eq:eq4}
\end{eqnarray}
where $GM$ is the product of the gravitational constant and the total mass of the attracting body, $R$ is the radius of the reference sphere, $n$ and $m$ are the degree and order of the expansion, respectively, $\overline{P}_{nm}$ are the fully normalized associated Legendre functions of the first kind, $\overline{c}_{nm}$ and $\overline{s}_{nm}$ are the dimensionless spherical harmonics coefficients.

\subsection{Spheroidal harmonics}
Spheroidal coordinates rely on the definition of a reference figure, i.e. a spheroid with specified orientation and eccentricity. The counterpart to the radial coordinate $r$ is the semi-axis of a spheroid confocal with the reference ellipse: the semi-minor axis $u$ in case of oblate spheroids and the semi-major axis $v$ for prolate spheroids. The angular components are again the longitude $\lambda$ and the reduced colatitudes $\vartheta^{\left(o\right)}$ and $\vartheta^{\left(p\right)}$, where the superscripts $o$ and $p$ are introduced to distinguish between \textbf{o}blate and \textbf{p}rolate coordinates.

The spheroidal harmonic expansion of the gravitational potential is given by \citep{Hobson2012}
\begin{eqnarray}
   V \left( u, \vartheta^{\left(o\right)}, \lambda \right) &= & \frac{GM}{a_1} \sum_{n=0}^{\infty} \sum_{m=0}^{n} \frac{ Q_{nm} \left( iu/\varepsilon \right) }{ Q_{nm} \left( ia_2/\varepsilon \right) } \overline{P}_{nm} \left( \cos \vartheta^{\left(o\right)} \right) \nonumber \\
   && \times \left[ \overline{c}_{nm}^{\left(o\right)} \cos m \lambda + \overline{s}_{nm}^{\left(o\right)} \sin m \lambda \right],
\label{eq:eq5}
\end{eqnarray}
and
\begin{eqnarray}
   V \left( v, \vartheta^{\left(p\right)}, \lambda \right) &= & \frac{GM}{a_1} \sum_{n=0}^{\infty} \sum_{m=0}^{n} \frac{ Q_{nm} \left( iv/\varepsilon \right) }{ Q_{nm} \left( ia_1/\varepsilon \right) } \overline{P}_{nm} \left( \cos \vartheta^{\left(p\right)} \right) \nonumber \\
   && \times \left[ \overline{c}_{nm}^{\left(p\right)} \cos m \lambda + \overline{s}_{nm}^{\left(p\right)} \sin m \lambda \right],
\label{eq:eq6}
\end{eqnarray}
where $a_1$ and $a_2$ are the semi-major and semi-minor axes of the reference spheroid, respectively, $\varepsilon$ is the linear eccentricity and $Q_{nm}$ are the associated Legendre functions of the second kind. For a thorough treatment of the theory of spheroidal harmonics we refer the reader to \citet{Byerly2003,Hobson2012}, effective algorithms for computing these functions can be found in \citet{Fukushima2013,Fukushima2014}.

\subsection{Ellipsoidal harmonics}
The tri-axial ellipsoidal coordinates are defined as \citep{Dassios2012}
\begin{eqnarray}
 \rho &= & \sqrt{a_1^2-q_1} \nonumber \\
 \mu  &= & \sqrt{a_1^2-q_2} \\ 
 \nu  &= & \sqrt{a_1^2-q_2}, \nonumber
 \label{eq:eq7}
\end{eqnarray}
where $q_1,q_2,q_3$ are the real roots of the cubic polynomial
\begin{equation}
 \frac{x^2}{a_1^2-q} + \frac{y^2}{a_2^2-q} + \frac{z^2}{a_3^2-q} = 1
 \label{eq:eq8}
\end{equation}
and $a_1,a_2,a_3$ are the descendingly ordered semi-axes of a reference ellipsoid centered at its origin.

The exterior potential parameterized in ellipsoidal harmonics is given by
\begin{eqnarray}
   V \left( \rho, \mu, \nu \right) &= & GM \sum_{n=0}^{\infty} \sum_{m=0}^{2n} \overline{\alpha}_{nm} \frac{ F_{nm} \left( \rho \right) }{ F_{nm} \left( a_1 \right) } \nonumber \\
   && \times \overline{E}_{nm} \left( \mu \right) \overline{E}_{nm} \left( \nu \right),
 \label{eq:eq9}
\end{eqnarray}
where $\overline{E}_{nm}$ and $F_{nm}$ are the Lam\'{e} functions of the first and the second kind, respectively. The second-kind function, $F_{nm} \left( \rho \right)$, accounts for the radial attenuation of the gravitational signal, analogous to the functions $Q_{nm}$ in the spheroidal case. The coefficients $\overline{\alpha}_{nm}$ correspond to the SH, OH and PH coefficients $\overline{c}_{nm}$ and $\overline{s}_{nm}$. 

\section{Ellipsoidal harmonics on the log-scale}
\label{sec:ehlog}
\subsection{Motivation}
\label{subsec:motivation}
Ellipsoidal harmonics are enormously laborious from a computational point of view \citep{Hu2012,Hu2015}. In contrast to the other parametrizations, no elegant recurrence formula is known that would enable a fast computation of the basis functions, i.e. the Lam\'{e} functions. Moreover, numerical issues arise when evaluating functions of higher orders, say beyond degree 15. Specifically, the determination of the normalization factor, referred to as $\gamma$ in the sequel, limits the applicability tremendously \citep{Bardhan2012}. This factor is needed to balance the large numerical quantities of the ellipsoidal surface harmonics (see eq.~\ref{eq:eq10}).

A further crucial aspect is the fact that the numerical values of the basis functions increase rapidly with growing expansion degree. As a consequence, arithmetic overflow might occur when directly evaluating the basis functions. For example, assuming a reference ellipsoid with semi-axes $a_1 =$~3~km, $a_2 =$~2~km, $a_3 =$~1~km, $\gamma$ is in the order of $10^{128}$ for degree 10 and $10^{256}$ for degree 20. The numerical values of the Lam\'{e} functions are similarly large. Programming environments which operate in accordance with the IEEE Standard 754 \citep{Zuras2008} are able to represent numbers of double-precision floating-point formats up to a maximum of almost $1.8 \times 10^{308}$. MATLAB~\textregistered, which was used for this work, belongs to this class of programs. 

We designed a series of tests to demonstrate the influence of overflow on the EH series expansions. First, we were interested in finding out how the shape of the reference ellipsoid, i.e. the two focal lengths, affects this issue. To achieve this, a set of reference ellipsoids with constant semi-major axis $a_1$ and variable semi-minor axes $a_2$ and $a_3$ was used. The values of the latter are controlled by the flattening parameters $f_{a_2} = \left( a_1 - a_2 \right) / a_1$ and $f_{a_3} = \left( a_1 - a_3 \right) / a_1$ with $0 < f_{a_2} < f_{a_3} < 1$. The gravitational potential, as given in eq.~\ref{eq:eq9}, was evaluated independently for each of these reference ellipsoids at eight uniformly distributed points on a circumscribed sphere. The $GM$ term was neglected. Starting at zero, the expansion degree of the series was successively increased until over- or underflow occurred. Since accuracy was not an issue here, a very simple midpoint approximation was used to accelerate the computation of the elliptical integrals appearing in the evaluation of the second-kind Lam\'{e} functions and $\gamma$ (see subsections~\ref{subsec:lame2} and \ref{subsec:norm} for details). 

\begin{figure}
\noindent\includegraphics[width=20pc]{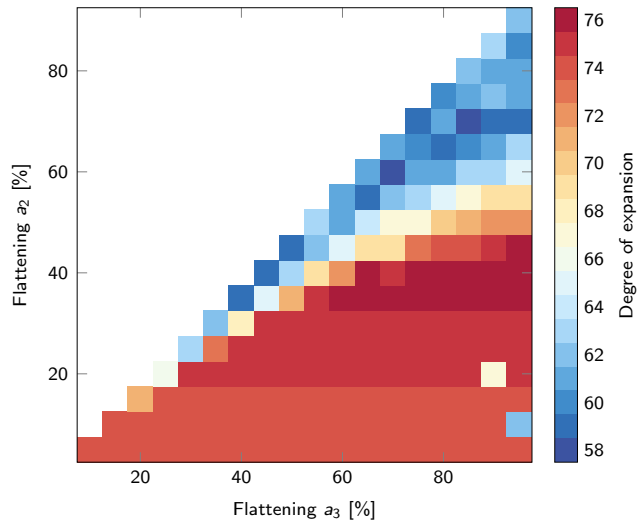}
\caption{Shape-dependent expansion limit of ellipsoidal harmonics series, exemplary for semi-major axis length $a_1 =$~100~m. Polynomials of higher degree cause arithmetic over- or underflow.}
\label{fig:fig1}
\end{figure}

Fig. \ref{fig:fig1} illustrates our findings for the exemplary case of $a_1 =$~100~m. Denoting $N$ as the truncation degree, the expansion limits range from $N=58$ to $N=76$ where the higher resolutions are obtained for ``oblate-like'' ellipsoids.

How does the \textit{size} of the ellipsoid affect the EH expansion? In order to investigate this question we repeated the previous test and extended our set of reference ellipsoids by letting the semi-major axis $a_1$ vary over several orders of magnitude. Fig. \ref{fig:fig2} shows the median value of each solution in dependence of the semi-major axis' length. We find a strong decline of the maximum obtainable resolution with increasingly large ellipsoids.

As a remedy for this problem, we suggest a reformulation of the various components of EH in terms of logarithmic expressions. Regarding cylindrical harmonics, i.e. solutions to Laplace's equation in cylindrical coordinates, similar investigations were made by \citet{Rothwell2005}. This parametrization is based upon the Bessel functions, which tend to over- or underflow for higher degrees as well. The author claims that the logarithmic approach is particularly useful when products or ratios of Bessel functions need to be determined. That is because the individual functions may over- or underflow, while the product of those is possibly representable.

Considering EH, the same train of thought can be followed. For instance, as explained in \citet{Dassios2012}, normalization (indicated by the vinculum) of the ellipsoidal surface harmonics is done via
\begin{equation}
  \overline{E} \left( \mu \right) \overline{E} \left( \nu \right) = \frac{ E \left( \mu \right) E \left( \nu \right)}{\sqrt{\gamma}}.
 \label{eq:eq10}
\end{equation}
This is a multiplication of two very large numbers followed by a division with another large number. Using the example from before, i.e. an ellipsoid with semi-axes $a_1 =$~3~km, $a_2 =$~2~km, $a_3 =$~1~km, evaluation of the degree 30-functions at some location on the surface yields values as large as $10^{190}$ for the numerator and $10^{384}$ for $\gamma$. While the latter itself clearly exceeds the maximum representable value, division with its square root would result in an easy to handle value of about $10^{-2}$. In the following subsections we demonstrate how this can be achieved when using logarithmic expressions.

\subsection{Logarithmic identities}
\label{subsec:identities}
For the convenience of the reader we first summarize the most important logarithmic rules and identities before diving into the details of the EH. For two positive real numbers $d_1$ and $d_2$ the following rules hold true \citep{Abramowitz1965}:
\begin{eqnarray}
   \log_b \left( b^p \right) &= &p, \label{eq:eq11}  \\ 
   b^{\log_b d_1} &= &d_1, \label{eq:eq12}  \\ 
   \log_b \left( d_1 d_2 \right) &= &\log_b d_1 + \log_b d_2, \label{eq:eq13}  \\  
   \log_b \left( d_1 / d_2 \right) &= &\log_b d_1 -\log_b d_2, \label{eq:eq14}  \\
   \log_b \left( d_1^p \right) &= &p \log_b d_1, \label{eq:eq15}. 
\end{eqnarray}
The base $b$ and the exponent $p$ are real numbers; the former must be positive. As an extension to the basic rules, summation and subtraction can be reformulated under the condition that $d_1 > d_2$ as
\begin{eqnarray}
   \log_b \left( d_1 + d_2 \right) &= &\log_b d_1 + \log_b \left( 1 + b^{ \log_b d_2 - \log_b d_1} \right) \label{eq:eq16}, \\
   \log_b \left( d_1 - d_2 \right) &= &\log_b d_1 + \log_b \left( 1 - b^{ \log_b d_2 - \log_b d_1} \right) \label{eq:eq17}.
\end{eqnarray} 
It is easily seen now that the logarithm of eq.~\ref{eq:eq10} simplifies to a sequence of simple arithmetic operations:
\begin{eqnarray}
  \log_b \left(\overline{E} \left( \mu \right) \overline{E} \left( \nu \right) \right) &= &\log_b E \left( \mu \right) + \log_b E \left( \nu \right) \nonumber \\
  &&- \frac{1}{2} \log_b \gamma.
 \label{eq:eq18}
\end{eqnarray}
Recalling the example in subsection~\ref{subsec:motivation} and setting $b=10$, we find $190 - 0.5 \times 384 = -2$ for the right-hand side of the eq.~\ref{eq:eq18}. Back-transformation is achieved using the identity in eq.~\ref{eq:eq12}.

\subsection{Lam\'{e} functions of the first kind}
The method for computing the Lam\'{e} functions was derived by \citet{Ritter1998}. It involves polynomials of the type
\begin{equation}
 T_{nm} \left( w_i \right) = \sum_{j=0}^{N_{\mathcal{F}} - 1} \kappa_j \left( 1 - \frac{w_i^2}{k_2^2} \right)^j,
 \label{eq:eq19}
\end{equation}
where $w_i$ is one of the three ellipsoidal coordinates, $k_2$ is the semi-focal length $k_2 = \sqrt{a_1^2-a_2^2}$ and $N_{\mathcal{F}}$ is the number of functions associated with one of the four solution classes $\mathcal{F}$ for a given degree. The polynomial coefficients $\kappa_j$ are obtained by eigenvalue-eigenvector-decomposition \citep{Dobner1998}. Based on the polynomials $T_{nm}$, an individual Lam\'{e} function is computed by multiplication with the coordinate-dependent quantity $\psi_{nm} \left( w_i \right)$:
\begin{equation}
 E_{nm} \left( w_i \right) = \psi_{nm} \left( w_i \right) T_{nm} \left( w_i \right).
 \label{eq:eq20}
\end{equation}
Speaking in terms of overflow issues, the computation of $\kappa_j$ and $\psi_{nm}$ is harmless and can be carried out in a straightforward manner. The sum in eq.~\ref{eq:eq19}, however, must be taken care of. Using the identities from before, we can rewrite the summands as
\begin{equation}
 \log_b \kappa_j + j \log_b \left( 1 - \frac{w_i^2}{k_2^2} \right).
 \label{eq:eq21}
\end{equation}
Most importantly, instead of raising the expression in the brackets to the power of $j$, the reformulation results in the multiplication with $j$. This fact alone increases the computable resolution tremendously.
The next crucial part is the actual summation of the individual terms. Before applying eq.~\ref{eq:eq16} we need to order the summands in a descending manner. The apparent problem here is that we deal with logarithms and not the actual values of the individual terms. If, however, the base is chosen in such way that $b > 1$ the inequality $d_1 > d_2$ also holds true on the log-scale, i.e. $\log d_1 > \log d_2$ \citep{Abramowitz1965}. 

The logarithm of the Lam\'{e} functions is then given by
\begin{equation}
 \log_b E_{nm} \left( w_i \right) = \log_b \psi_{nm} \left( w_i \right) + \log_b T_{nm} \left( w_i \right).
 \label{eq:eq22}
\end{equation}

\subsection{Lam\'{e} functions of the second kind}
\label{subsec:lame2}
The functions of the second kind can be computed by
\begin{equation}
 F_{nm} \left( \rho \right) = E_{nm} \left( \rho \right) I_{nm} \left( \rho \right)
 \label{eq:eq23}
\end{equation}
where $I_{nm}$ are integrals of the form
\begin{equation}
 I_{nm} \left( \rho \right) = \int_{0}^{\rho^{-1}} \frac{t^{2n} dt}{ \left( E_{nm}\left( t \right) \right)^2 \sqrt{1-k_3^2 t^2} \sqrt{1-k_2^2t^2}},
 \label{eq:eq24}
\end{equation}
with the second focal length $k_3 = \sqrt{a_1^2-a_3^2}$. The integral is usually solved by means of numerical quadrature. Basically, this is nothing but a (weighted) sum of function values. For example, consider the most trivial quadrature method, the midpoint rule, to approximate the definite integral of some function $g\left(t\right)$ in the interval $\left[p,q\right]$ with $l$ subintervals \citep{Suli2003}:
\begin{equation}
 \int_p^q g \left( t \right) dt \approx \frac{q-p}{l} \sum_{j=0}^{l-1} g \left( p + \frac{q-p}{2l} + j \frac{q-p}{l} \right).
 \label{eq:eq25}
\end{equation}
Assuming $g \left(t\right)$ is the integrand in eq.~\ref{eq:eq24} then its logarithm is given by
\begin{eqnarray}
 \log_b g\left( t \right) &=& 2n \log_b t - 2 \log_b E_{nm}\left( t \right) \nonumber\\
 &&- \frac{1}{2} \left( \log_b \left( 1-k_3^2 t^2 \right) + \log_b \left( 1-k_2^2 t^2 \right) \right)
 \label{eq:eq26}
\end{eqnarray}
and determination of $I_{nm}$ on the log-scale is achieved using the identities eq.~\ref{eq:eq16} and eq.~\ref{eq:eq17}. 

The logarithm of the Lam\'{e} functions $F_{nm} \left( \rho \right)$ is obtained as
\begin{equation}
  \log_b F_{nm} \left( \rho \right) = \log_b E_{nm} \left( \rho \right) + \log_b I_{nm} \left( \rho \right).
 \label{eq:eq27}
\end{equation}

\subsection{Normalization constant}
\label{subsec:norm}
The normalization formula reads
\begin{equation}
 \gamma_{nm} = 4\pi \left( \alpha B - \beta A \right),
 \label{eq:eq28}
\end{equation}
where $\alpha, \beta, A, B$ are solutions of a system of equations involving four elliptic integrals. For instance, the solution for $\alpha$ in the explicit form is
\begin{equation}
  \alpha =  \frac{\mathcal{I}_1 I_3^1 - \mathcal{I}_3 I_2^1}{I_2^0 I_3^1 - I_2^1 I_3^0},
 \label{eq:eq29}
\end{equation}
where $\mathcal{I}_i$ are elliptic integrals, which can be expressed as a linear combination of basic integrals $I_j^k$ (details and notation see \citet{Garmier2001}). The integrals themselves as well as the quotient in the above equation are computed logarithmically. To clarify this process we state the individual steps in more detail. We first simplify the notation to
\newline
\begin{eqnarray}
 s_1 &=& \mathcal{I}_1 I_3^1, \\
 s_2 &=& \mathcal{I}_3 I_2^1, \\
 s_3 &=& I_2^0 I_3^1, \\
 s_4 &=& I_2^1 I_3^0
 \label{eq:eq30}
\end{eqnarray}
and find the corresponding logarithms to be
\begin{eqnarray}
 \log_b s_1 &=& \log_b \mathcal{I}_1 + \log_b I_3^1, \\
 \log_b s_2 &=& \log_b \mathcal{I}_3 + \log_b I_2^1, \\
 \log_b s_3 &=& \log_b I_2^0 + \log_b I_3^1, \\
 \log_b s_4 &=& \log_b I_2^1 + \log_b I_3^0.
 \label{eq:eq31}
\end{eqnarray}
Using the quotient and the subtraction rule and assuming that $s_1 > s_2$ and $s_3 > s_4$ we find
\begin{eqnarray}
 \log_b \alpha &=& \log_b \left( s_1 - s_2 \right) - \log_b \left( s_3 - s_4 \right) \nonumber\\
 &=& \left[ \log_b s_1 + \log_b \left( 1 - b^{\log_b s_2 - \log_b s_1} \right) \right] \nonumber\\
 &&- \left[ \log_b s_3 + \log_b \left( 1 - b^{\log_b s_4 - \log_b s_3} \right) \right].
 \label{eq:eq32}
\end{eqnarray}
If the inequalities postulated before do not hold true, the subtraction identity must be adapted accordingly. The other three constants $\beta, A, B$ are obtained similarly. 

Finally, the logarithm of the normalization factor is computed via
\begin{eqnarray}
 \log_b \gamma_{nm} &=& \log_b \left(4\pi\right) + \log_b \left( \alpha B - \beta A \right) \nonumber\\
 &=& \log_b \left(4\pi\right) + \log_b \left( \alpha B \right) \nonumber\\
 &&+ \log_b \left( 1 - b^{ \log_b\left(\beta A \right) - \log_b\left(\alpha B\right) } \right)
 \label{eq:eq39}
\end{eqnarray}
with $\log_b\left(\alpha B\right) = \log_b \alpha + \log_b B$ and $\log_b\left(\beta A\right) = \log_b \beta + \log_b A$. Again, according changes must be made if $\beta A > \alpha B$.

\subsection{Putting it all together}
In eq.~\ref{eq:eq9} it was shown that the gravitational potential in the ellipsoidal harmonic parameterization involves Lam\'{e} functions of the first and second kind. Computation of these functions can be carried out on the basis of the logarithmic identities presented in subsection~\ref{subsec:identities}. The logarithmic expressions for the Lam\'{e} functions of the first kind, $E_{nm}$, and for those of the second kind, $F_{nm}$, as well as for the underlying normalization factor $\gamma_{nm}$ are stated in eqs.~\ref{eq:eq22}, \ref{eq:eq27} and \ref{eq:eq39}, respectively. Under the consideration of these definitions the gravitational potential expansion can be written as
\begin{equation}
  V  = GM \sum_{n=0}^{\infty} \sum_{m=0}^{2n} \overline{\alpha}_{nm} \times b^{\log_b L_{nm}}
 \label{eq:eq40}
\end{equation}
with $L_{nm}$ being a shorthand notation for the product of the basis functions:
\begin{eqnarray}
   L_{nm} &=& \frac{I_{nm} \left( \rho \right)}{I_{nm} \left( a_1 \right)} \frac{1}{\sqrt{\gamma_{nm}}} \nonumber\\ 
   &&\times\frac{ \psi_{nm} \left( \rho \right) \psi_{nm} \left( \mu \right) \psi_{nm} \left( \nu \right) }{\psi_{nm} \left( a_1 \right)} \nonumber\\
   &&\times\frac{ T_{nm} \left( \rho \right) T_{nm} \left( \mu \right) T_{nm} \left( \nu \right) }{T_{nm} \left( a_1 \right)}.
 \label{eq:eq41}
\end{eqnarray}
According to \citet{Garmier2001}, the triple product of the functions $\psi_{nm} \left( w_i \right)$, denoted by the capital letter $\Psi_{nm} \left( x, y, z \right)$, can be expressed in terms of Cartesian coordinates (Table~3, ibid.). This is necessary in order to avoid sign ambiguities. 
Furthermore, it is worth pointing out that the coefficients $\kappa_j$ occurring in eq.~\ref{eq:eq19} are independent of the coordinate $w_i$ and, thus, need only be once computed for the reference ellipsoid. This accelerates the computation of the last quotient in eq.~\ref{eq:eq41}.

The logarithmic equivalent of $L_{nm}$ can be written as
\begin{eqnarray}
 \log_b L_{nm} &=& \log_b I_{nm} \left( \rho \right) + \log_b \Psi_{nm} \left( x, y, z \right) \nonumber\\
 &&+ \log_b T_{nm} \left( \rho \right) + \log_b T_{nm} \left( \mu \right) + \log_b T_{nm} \left( \nu \right) \nonumber\\
 &&- \log_b I_{nm} \left( a_1 \right) - \frac{1}{2} \log_b \gamma_{nm} - \log_b \psi_{nm} \left( a_1 \right) \nonumber\\
 &&- \log_b T_{nm} \left( a_1 \right) .
 \label{eq:eq42}
\end{eqnarray}
This rather cumbersome approach has the advantage of allowing the computation of very high degree harmonics (we tested up to $N=500$) without the issue of overflow. Of course, the increased number of arithmetic operations with this method results in higher computation costs.

\subsection{Some important considerations}
\subsubsection{Choosing the base}
In order to apply the summation and subtraction identities the base of the logarithm must be greater than $1$. We used $b=10$ in our tests, but any other real number fulfilling this condition is fine.

\subsubsection{Computation on the Cartesian planes}
The logarithm of zero is undefined, i.e. \citep{Abramowitz1965}
\begin{equation}
 \lim\limits_{d_1 \to 0^+} \log_b d_1 = -\infty.
 \label{eq:eq43}
\end{equation}
This issue will arise when computing the ellipsoidal harmonics on the Cartesian planes. That is, because at these positions, at least one of the ellipsoidal angular coordinates $\mu$ or $\nu$ is equal to either one of the semi-focal lengths $k_2$ or $k_3$; as a consequence, expressions like the one in eq.~\ref{eq:eq21} become zero \citep[pp. 8-13]{Dassios2012}. In order to obtain real numbers, the affected Lam\'{e} functions must be excluded from the logarithmic algorithm and set to zero after back-transformation is completed (eq.~\ref{eq:eq40}).

\subsubsection{Dealing with negative values}
In the real number system the logarithm of a negative number is not defined. Instead complex numbers are used. However, it is an easy task to separate the signs before computing the logarithm of the absolute values and restore them after the computation is done. Of course, this means that computation on the linear scale must be carried out for the signs as well which results in more computational effort. For instance, consider the multiplication of $d_1 =-4$ and $d_2 =32$ with the result $d_1 d_2=-128$ on the log-scale with base 2:
\begin{eqnarray}
 \textrm{sgn} \left(d_1 d_2\right) &=& \textrm{sgn} \left(d_1 \right) \textrm{sgn} \left(d_2 \right) = -1 \times 1 = -1 \nonumber\\
 \log_{2} \left(\left|d_1 \right| \left|d_2 \right|\right) &=& \log_{2} \left|d_1 \right| + \log_{2} \left|d_2 \right| = 2 + 5 = 7 \nonumber\\
 d_1 d_2  &=& \textrm{sgn} \left(d_1 d_2 \right) \times 2^{\log_{2} \left(\left|d_1 \right| \left|d_2 \right|\right)} \nonumber\\
 &=& -1 \times 2^7 = -128.
 \label{eq:eq44}
\end{eqnarray}
This is slightly more difficult when dealing with sums. Consider any two real numbers $d_1 $ and $d_2 $ with arbitrary signs. We introduce the vector $\mathbf{\tau}$, which comprises the descendingly sorted absolute values of those two numbers, and the vector $\mathbf{\sigma}$ containing the corresponding signs. The logarithm of the absolute value of the summation, i.e. $\log_b \left|d_1 +d_2 \right|$, is then achieved by distinguishing between the cases
\begin{eqnarray}
 \log_b \tau_1 &+& 
 \cases{ 
 \log_b \left( 1 + b^{\log_b \tau_2 - \log_b \tau_1} \right ) 
 &if \(\sigma_1 \times \sigma_2 = 1 \)\cr
 \log_b \left( 1 - b^{\log_b \tau_2 - \log_b \tau_1} \right) 
 &if \( \sigma_1 \times \sigma_2 = -1 \)\cr} 
 \label{eq:eq45}
\end{eqnarray}
The notation $\tau_i$ and $\sigma_i$ indicates the $i$th element of the vectors $\mathbf{\tau}$ and $\mathbf{\sigma}$. Finally, the actual value of the sum with the appropriate sign is obtained by
\begin{equation}
 d_1 +d_2  = \sigma_1 \times b^{\log_b \left|d_1 +d_2 \right|}.
 \label{eq:eq46}
\end{equation}
Demonstrating this procedure by the example $d_1 =-8$ and $d_2 =16$ we find $\mathbf{\tau}=\left[16, 8\right]$, $\mathbf{\sigma}=\left[1, -1\right]$ so that
\begin{equation}
 \log_2 \left|d_1 +d_2 \right| = \log_2 16 + \log_2 \left( 1 - 2^{\log_2 8 - \log_2 16} \right)
 \label{eq:eq47}
\end{equation}
and
\begin{equation}
 d_1 +d_2  = \sigma_1 \times 2^{\log_2 \left| d_1 +d_2  \right|} = 1 \times 2^3 = 8.
 \label{eq:eq48}
\end{equation}
When implementing this approach in a computer program, of course more than two summands can be dealt with at once.

\section{Comparison of SH, OH, PH and EH using the logarithm method}
\subsection{Method}
\label{sec:method}
We conducted a series of simulation tests in order to assess the accuracy and applicability of the spherical, spheroidal and ellipsoidal gravitational field parametrizations. Based on the polyhedral shape model of a Small Solar System Body we estimated in a first step the (axes-aligned) radii of the respective reference figures. This task was carried out in a quasi-random manner, meaning that the parameters of the minimum volume enclosing sphere, spheroid and ellipsoid were approximated iteratively using random numbers within a predefined range. While this is far from being an optimal solution we still claim that for our purpose the discrepancy between these ``random surfaces'' and the actual Brillouin surfaces is secondary and does not influence the conclusions of our study. In fact, the problem of computing the minimum volume enclosing ellipsoid of a set of data points is still an active field of research \citep[e.g.,][]{Todd2007,Kumar2008,Ahipasaoglu2015}.

Next we used the forward-modeling technique presented in \citet{Werner1994} to determine the true gravitational field of the object. This is done under the assumption of constant mean bulk density. Based on the Reuter grid algorithm \citep{Reuter1982} we evenly distributed the evaluation points on the surface of a sphere enclosing the aforementioned Brillouin surfaces and, of course, the polyhedron itself. Compared to the geographical grid, the point density is loosened due to the equi-distant characteristic of the Reuter grid, especially near the poles. In addition to the gravitational potential also its first derivative, i.e. the gravitational accelerations were computed.

The most commonly used method for determining the unknown coefficients of the harmonic expansions is to apply the orthogonality relations of the basis functions and to integrate over the respective reference surface \citep{Hofmann-Wellenhof2006}. In this work we make use of a different approach: the least squares adjustment. Regarding the potential evaluations as ``observations'' we can set up a functional model which relates the simulated values (potential or accelerations) to harmonic coefficients. This way the backward-modeling problem reduces to solving a system of linear equations. Another advantage of this method lies in the possibility to account for uncertainties (e.g. shape model errors) when including a stochastic model. However, this topic exceeds the scope of this paper. 

The process of using the respective coefficients to compute the gravitational quantities is referred to as synthesis. The differences between the analytically determined values and those resulting from the synthesis allow for the interpretation of the performance of the chosen parametrization. This misfit is introduced as the percentage error $\delta_{\left( V,g \right)}$, defined as
\begin{equation}
 \delta_V = \frac{V_b - V_f}{V_f} \times 100, \quad \delta_g = \frac{\|\mathbf{g}_b - \mathbf{g}_f\|}{\| \mathbf{g}_f\|} \times 100.
 \label{eq:eq49}
\end{equation}
The subscripts $f$ and $b$ denote forward and backward, respectively. Note that $\delta_g$ is a scalar and represents the error of the magnitude of the acceleration vector $\mathbf{g}_b$. $V$, as usual, is the potential.

The definition of the Brillouin surfaces and the issue of divergence of the harmonic series was introduced in section~\ref{sec:intro}. We were interested in assessing the effects of possible divergence in our simulations. Therefore, the forward-calculation step involving the polyhedral gravitation method was repeated for a regular grid of points on the surface of the body. Next, the harmonic synthesis of the gravitational field functionals was carried out at these surface locations based on the SH, OH, PH and EH coefficients obtained from the previous simulation (i.e. from observations on the circumscribed sphere). The differences between the simulated values and the respective approximations were again quantified by means of eq.~\ref{eq:eq49}.

\subsection{Example: Comet 67P/Churyumov-Gerasimenko}
We used an early version of the shape model developed by the mission teams consisting of 62908 faces \citep{Preusker2015}. Important physical parameters including estimations for the nucleus' volume, mass and density were taken from \citet{Sierks2015}. The shape model was rescaled to match the real volume based on the theory in \citet{Newson1899}. Table~\ref{tab:tab2} summarizes the physical parameters of 67P and states our approximations of the radii of the Brillouin sphere, spheroids (oblate and prolate) and ellipsoid. A visual comparison of the reference sphere, spheroids and ellipsoid is given in the left panel of Fig.~\ref{fig:fig3}. To get a better understanding of their appropriateness this figure also features a simplified version of the shape model consisting of 1000 faces.

\subsubsection{Harmonic analysis/synthesis outside the Brillouin surfaces}
The true gravitational field was evaluated for 7124 points on the surface of a sphere of radius $R =$~3000~m. This corresponds to the Reuter grid resolution of 75 meridional points. In the right panel of Fig.~\ref{fig:fig3} we estimated the true potential by means of the SH, OH, PH and EH up to degree 10 and assessed the quality by means of the percentage error $\delta_V$. We found good convergence of each of the series with an average accuracy of better than 1~\% in all cases, see statistics in Table~\ref{tab:tab3}. The largest errors occur in close proximity to the small lobe of the comet because in this area the signal is not attenuated as much as elsewhere. The characteristics of the error patterns mirror the geometrical suitability of the respective parametrizations. This is particularly well visible when comparing the SH and OH cases. Compared to the sphere, the oblate reference spheroid fits the comet much better near the poles which, as a consequence, causes a decrease of the modeling errors at these latitudes. All in all the prolate spheroid and the ellipsoid approximate the nucleus' shape best resulting again in an overall decrease of the associated errors.

How do higher harmonic degrees affect the accuracy of the various solutions? In order to answer this question we applied the algorithm presented in section~\ref{sec:ehlog} to compute EH on the log-scale. We expanded the gravitational field up until degree $N=40$. In Fig.~\ref{fig:fig4} the root mean square (rms) of the errors $\delta_V$ and $\delta_g$ are displayed as a function of the series expansion degree. Solid lines indicate that least squares adjustment was based on potential values, dashed lines on accelerations. 

All curves converge steadily towards the true potential. However, convergence takes place at noticeable faster pace for the PH and EH series. In fact, these two parametrizations perform almost equally well, particularly the harmonics of lower degree (e.g. $N\le15$). 

The errors associated with accelerations are consistently larger throughout. This systematic offset is explained by the fact that differentiating a function in the time (or spatial) domain means emphasizing the higher frequencies in the frequency domain \citep{Abramowitz1965a}. As a consequence, a much larger number of polynomial terms is needed in order to attain the same accuracy as for potential observations. For instance, the degree-10 potential field is just as accurate as the degree-20 acceleration field. Similar conclusions were drawn in \citet{Hu2015}; apart from the error in the magnitude of the vector, the authors also analyzed the model errors in the direction of the acceleration which they expressed in terms of the gravitational slope.

\subsubsection{Harmonic synthesis inside the Brillouin surfaces}
\label{subsubsec:churydiv}
We determined the gravitational effects at the centroids of the polygonal faces making up the polyhedron. Based on the results from the analysis step in the previous subsection, we synthesized the gravitational field on the surface locations in terms of SH, OH, PH and EH. The associated percentage errors $\delta_V$ are shown in Fig.~\ref{fig:fig5}. To facilitate the visual comparison, we scaled the color axis to a minimum and maximum of $\pm 10$~\%. However, the largest errors hugely exceed these limits. 
The concave shape of the comet causes a large volume of divergence, i.e. empty space between the topography and the Brillouin surfaces. This effect can clearly be seen in Fig.~\ref{fig:fig5}. The striped pattern of the spherical and spheroidal errors, which is already known from Fig.~\ref{fig:fig3} and caused by the neglect of higher degrees, is significantly amplified on this so-called neck region of the comet. The apparent change of the direction of the errors associated with the PH is somewhat misleading. It must be noted here, that in the prolate spheroidal coordinate system, the semi-major axis is aligned with $z$-axis, thus, the error stripes run from pole to pole - just as with the SH and the OH. None of the spherical and spheroidal parameterizations yields tolerable results in that region; the overall rms values of the SH, OH and PH approximations are $5.74 \times 10^{4}$, $4.33 \times 10^{1}$ and $1.85 \times 10^{3}$, respectively. At these polar regions, the OH are superior to the PH. The corresponding maximum errors, however, range up to 600~\% and relativize this superiority.
On the other hand, the EH synthesis could approximate the true gravitational field comparably well. The percentage errors range from -14~\% to 5~\% with a rms value of $2.67$~\%. The geometrical misfit seems to be better handled with this parameterization.

\subsection{Example: Objects from the DAMIT database}
\label{subsec:damit}
Our results related to comet 67P came at some surprise and have not been expected like this. The fact that the prolate spheroidal parametrization is virtually just as accurate as the ellipsoidal one provided the impulse for us to pose a follow-on question: Is it even necessary to use ellipsoidal harmonics? If not, the laborious procedure of computing the Lam\'{e} functions could be avoided by making use of the much simpler Legendre functions. 

To investigate this question thoroughly we applied the simulation technique outlined in subsection~\ref{sec:method} to an extended data set of solar system bodies. An excellent collection of asteroid shape models is given by the Database of Asteroid Models from Inversion Techniques, short DAMIT \citep{Durech2010}. Most of the polyhedrons have calibrated size, i.e. they are scaled to the actual physical dimensions of the body. However, some are unit sized. To get as much data as possible, we rescaled every asteroid shape model in such way that the final volume was equal to unity. Accordingly, the mean bulk density was set to unity for all objects. The shape models were rotated about the third axis to make sure that the maximum equatorial radius is aligned with the prime meridian. In order to improve the geometrical fitting of the Brillouin spheres, spheroids and ellipsoids to the polyhedra, we included the estimation of a translation vector in the algorithm for finding the reference surface parameters.

A complete list of asteroids involved in this study as well as numerical results of the experiments conducted in the following subsections is available in the supporting information of this article.

\subsubsection{Geometrical study of the samples}
First, we conducted a statistical analysis of the shapes of the asteroids to emphasize the fact that most Small Solar System Bodies are in fact irregular in shape. To this end, we approximated the minimum bounding boxes of the point sets using the algorithm in \citet{Vecchio2012} and analyzed, how much the body deviates from the ideal shape of a sphere. We introduce the shape measure $K_s$ as an index of spheroidicity to assess this characteristic,
\begin{equation}
 K_s = \left(1 - \frac{area_{l_2,l_3}}{area_{l_1,l_2}}\right) \times 100 = \left(1 - \frac{l_3}{l_1}\right) \times 100 ,
 \label{eq:eq50}
\end{equation}
with $l_1$, $l_2$, and $l_3$ being the descendingly sorted side lengths of the bounding box and $area_{l_1,l_2}$ and $area_{l_2,l_3}$ the areas of the corresponding faces. A perfect sphere has the index $K_s=0$~\%, the extremum of $K_s=100$~\% would either imply a flat circle (if oblate) or a straight line (if prolate). The histogram in Fig.~\ref{fig:fig6} reveals that the majority of the analyzed samples are moderately spheroidal, i.e. have indices ranging between 30 and 50~\%. Only a few bodies are almost perfectly spherical in shape (with indices less than 10~\%) and none of the tested samples exceeds 80~\% spheroidicity. We therefore conclude, that the use of OH or PH and EH should be considered as an alternative to SH in most of the cases.

The index $K_s$ is a useful measure of spheroidicity, however, it does not distinguish between oblate and prolate spheroids. Therefore, we try to answer the question of oblateness or prolateness by means of the volumes of the Brillouin spheroids. In subsection~\ref{subsubsec:churydiv}, we introduced the term divergence volume as the empty space inside the Brillouin surface. We express this misfit in terms of the percentage factor $K_V$:
\begin{equation}
 K_V = \left(1 - \frac{volume_{polyhedron}}{volume_{Brillouin}}\right) \times 100.
 \label{eq:eq51}
\end{equation}
A perfect fit of the reference surface to the polyhedron is obtained if $K_V = 0$~\%. The analysis of the samples from the database revealed average values of 59~\% for the Brillouin spheres, 47~\% and 45~\% for the oblate and prolate Brillouin spheroids, respectively, and 39~\% for the Brillouin ellipsoids (Fig.~\ref{fig:fig7}). As expected, the tri-axial ellipsoids are in the mean the most appropriate reference figures, however, the two types of spheroids are close seconds and differ by only 2~\% from each each other. 

\subsubsection{Comparison of the gravitational field solutions}
We chose the grid resolution to include 50 points along the meridians, yielding in total 3153 observations. Only potential values have been considered. The radius of this evaluation sphere was chosen to be in the same ratio to the reference surfaces as for comet 67P. The harmonics were estimated up to degree $N=25$ and evaluated in the synthesis step up to degree $N=10$. 

Here again, we were eager to analyze the effect of possible divergence on the surface of the bodies. Since many of the available shape models are tessellated in an irregular pattern with a range of differently shaped and sized triangles, the method of selecting the respective polygon centroids, which was applied in the case of comet 67P, would have led to a non-uniform evaluation point distribution. Instead, the Reuter grid defined on the circumscribed sphere was projected radially onto the surface of the model. 

To get a qualitative comparison between the spherical, spheroidal and ellipsoidal solutions we used the relative differences of the rms values of the respective simulation results. Expressed in terms of a formula this simply gives
\begin{equation}
 \Delta \delta_V = \frac{\textrm{rms } \delta_V^{\left(SH,OH,PH\right)} - \textrm{rms } \delta_V^{EH}}{\textrm{rms } \delta_V^{EH}}
 \label{eq:eq52}
\end{equation}

Figures~\ref{fig:fig8} and \ref{fig:fig9} show the results of our investigations. In Fig.~\ref{fig:fig8}, the gray dots indicate relative differences associated with either OH or PH solution, whichever is best. The black dots show the corresponding differences in SH. The results are visualized in dependence of the spheroidicity factor $K_s$ on the abscissa. The left panel refers to the investigation on the circumscribed sphere, the right panel to the divergence study on the surface. The bar chart in Fig.~\ref{fig:fig9} shows the mean values of the differences in intervals of ten. 

The visualizations in the respective figures prove clearly the aggravation of the spherical solutions with increasingly irregularly shaped bodies. Regarding the divergence study, this trend peaks dramatically in errors of over fifty thousand percent. The interpretation of the spheroidal solutions is not quite as straightforward. The analysis on the sphere reveals that almost spherical bodies with less than 20~\% spheroidicity seem to be approximated even better with OH or PH, as indicated by the negative sign. Though a flat ascending trend is visible here too, the average accuracy remains below 1~\% for all bodies. No obvious systematic trend can be observed in case of the divergence issue. The spheroidal and ellipsoidal solutions differ in the mean by only 1~\%, surprisingly, slightly larger errors are generated by EH if the shape of the asteroids exceeds 50~\% spheroidicity. 

\section{Conclusions}
Knowledge about modeling the Earth's gravity field has been used extensively over the last decades to describe the gravitational effects of celestial bodies. However, the increase of both effort and expenses put into space mission planning and operation demanded for more sophisticated techniques to attain the most possible accuracy.  Apart from navigational applications, this is particularly true for geophysical investigations. One of the more advanced methodologies is the parametrization in ellipsoidal harmonics. So far, the computation of these harmonics was limited to the low degrees due to numerical issues.

\begin{figure}
\noindent\includegraphics[width=20pc]{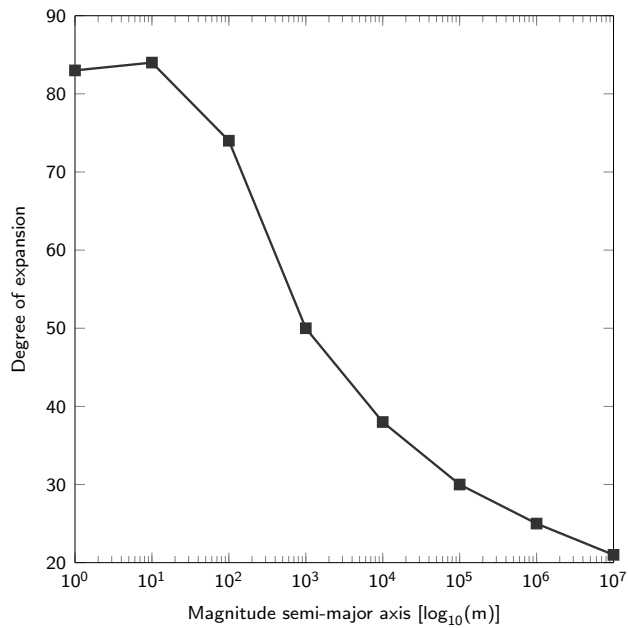}
\caption{Generalization of Fig. \ref{fig:fig1}. Median value of expansion limit of ellipsoidal harmonics series for differently shaped and sized reference ellipsoids.}
\label{fig:fig2}
\end{figure}

\begin{figure}
\noindent\includegraphics[width=40pc]{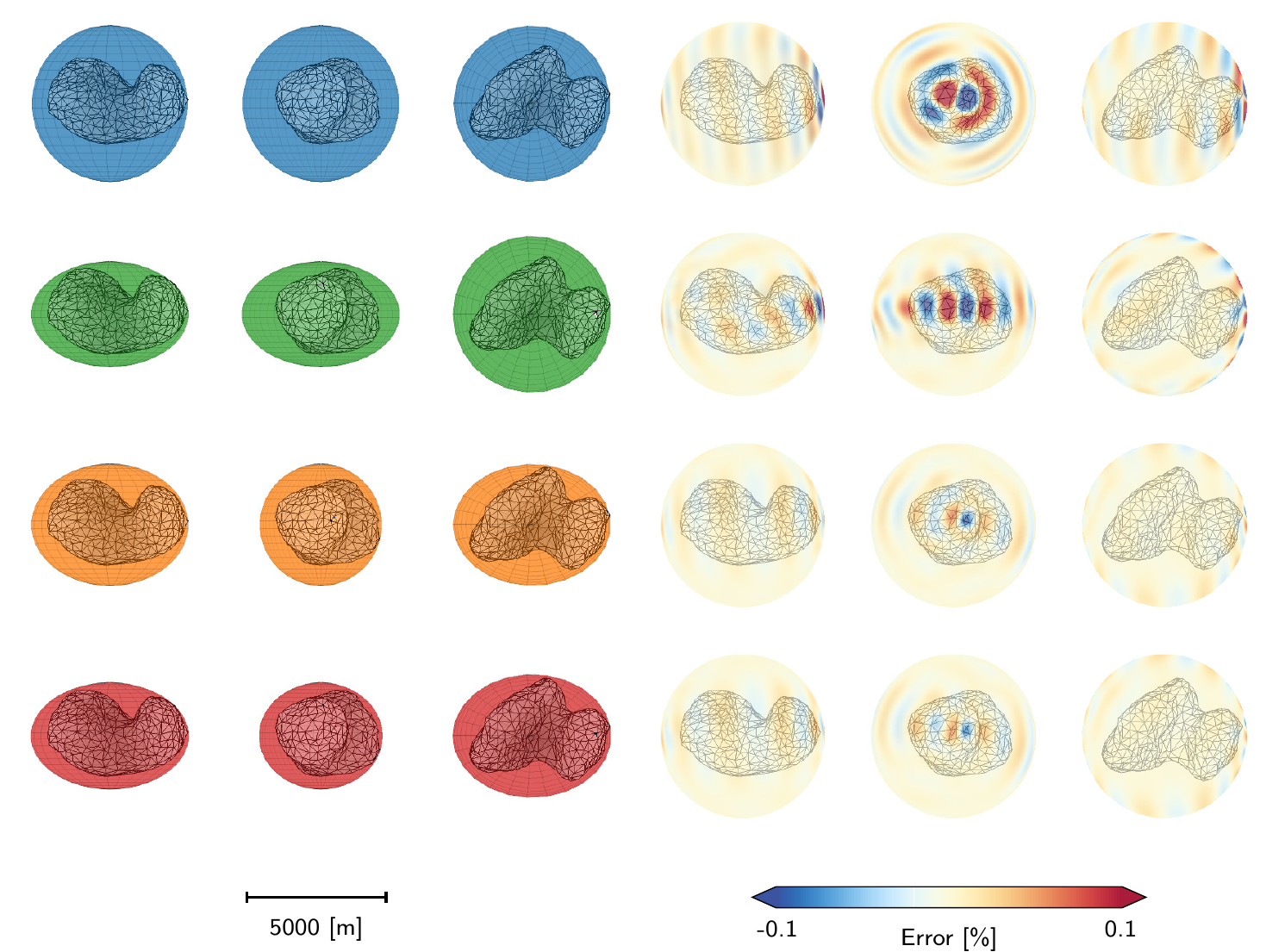}
\caption{Polyhedron model, Brillouin figures and simulation results for comet 67P/Churyumov-Gerasimenko. Left panel: Three normal views of the Brillouin sphere, oblate spheroid, prolate spheroid and ellipsoid (from top to bottom). Right panel: Three normal views of the percentage errors $\delta V$ on a circumscribed sphere with radius $R = 3000$~m, exemplary for an expansion up to degree $N = 10$. The approximation is according to spherical, oblate spheroidal, prolate spheroidal and ellipsoidal harmonics (from top to bottom).}
\label{fig:fig3}
\end{figure}

\begin{figure}
\noindent\includegraphics[width=20pc]{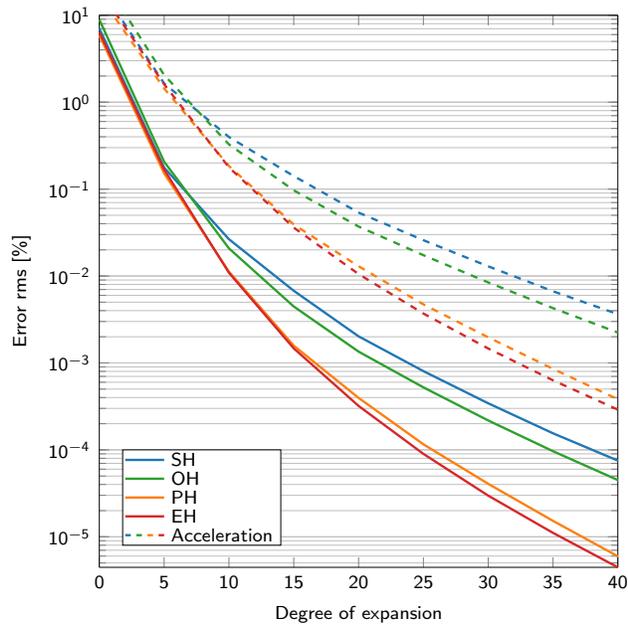}
\caption{Generalization of Fig. \ref{fig:fig3}, right panel. The rms values of the percentage errors $\delta_V$ are visualized in dependence of the expansion degree $N$. Solid lines indicate gravitational potential values, dashed lines gravitational accelerations.}
\label{fig:fig4}
\end{figure}

\begin{figure}
\noindent\includegraphics[width=20pc]{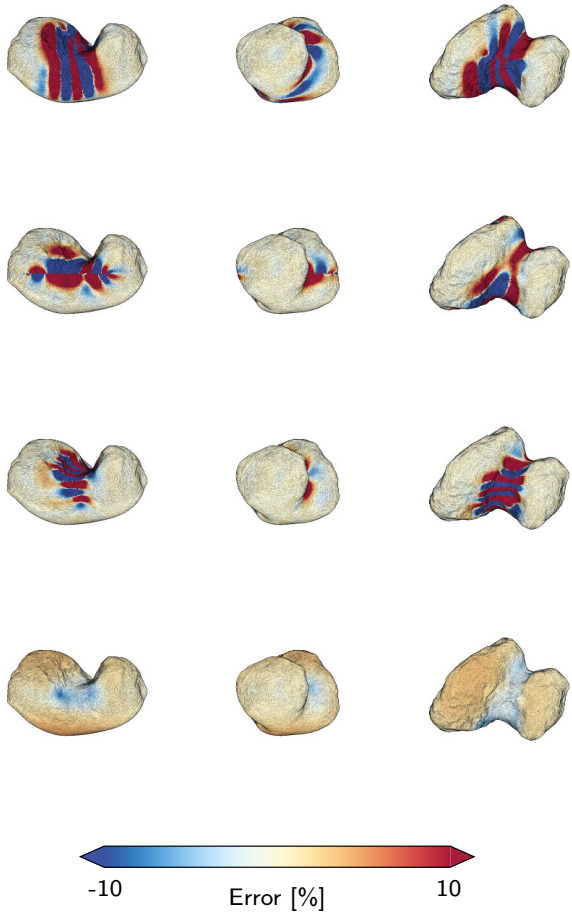}
\caption{Analogous setup as Fig. \ref{fig:fig3}, right panel. Here, however, the percentage errors $\delta V$ refer to evaluations on the surface of the comet.}
\label{fig:fig5}
\end{figure}

\begin{figure}
\noindent\includegraphics[width=20pc]{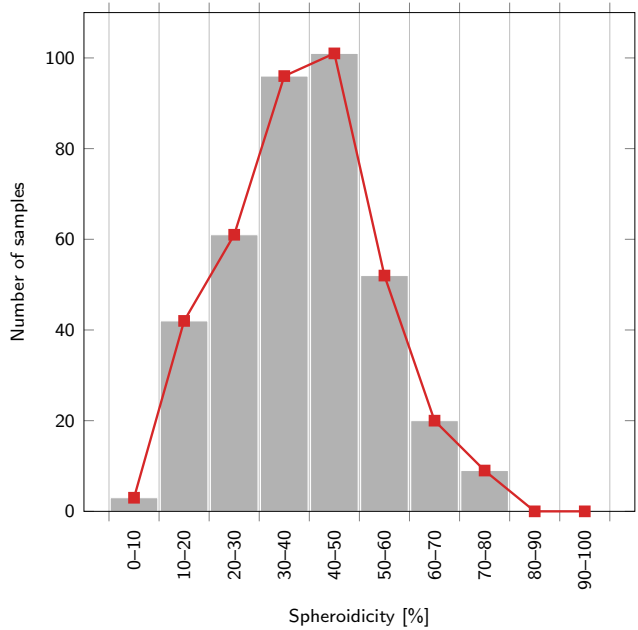}
\caption{Distribution of the index of spheroidicity $K_s$ for 384 samples from the DAMIT database. Note that a perfect sphere has the index $K_s=0$.}
\label{fig:fig6}
\end{figure}

\begin{figure}
\noindent\includegraphics[width=20pc]{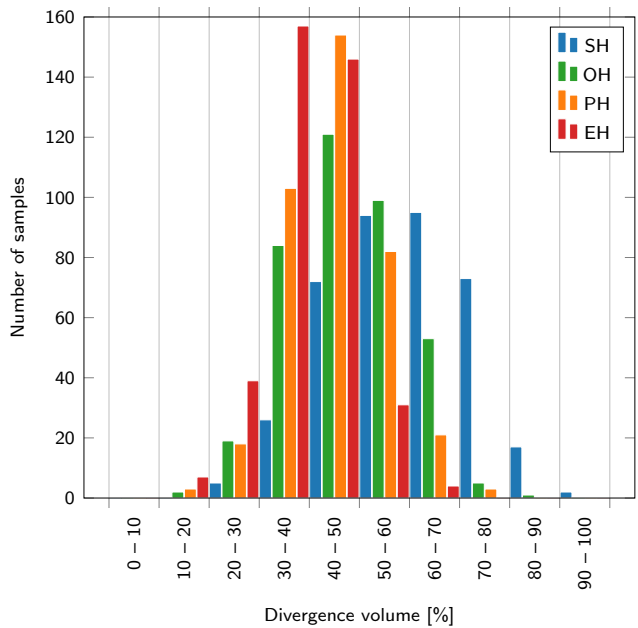}
\caption{Distribution of the divergence volume factor $K_V$ for 384 samples from the DAMIT database. Note that a perfect fit of the Brillouin surface to the body implies $K_V=0$. The sphere, the oblate spheroid, the prolate spheroid and the ellipsoid are represented by the colors blue, green, orange and red, respectively.}
\label{fig:fig7}
\end{figure}

\begin{figure}
\noindent\includegraphics[width=40pc]{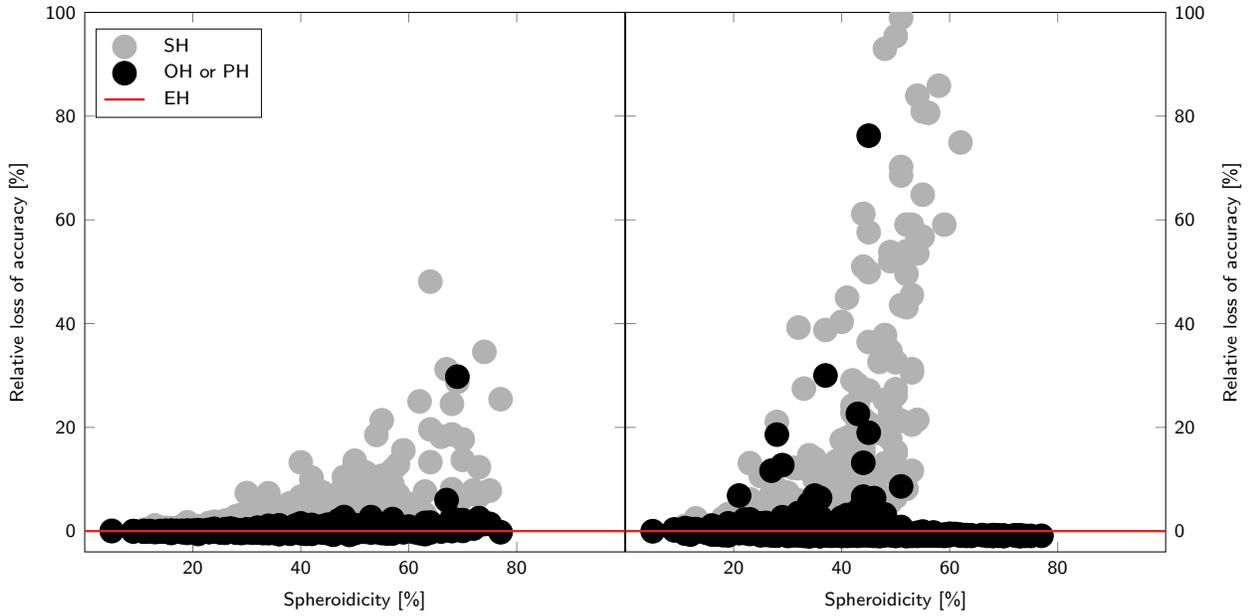}
\caption{Comparison of spherical, spheroidal and ellipsoidal solutions. Results are expressed as relative differences of the rms values of the percentage error $\delta V$ according to an expansion up to degree $N = 10$ for 384 celestial bodies from the DAMIT database. Left panel: analysis/synthesis on a circumscribed sphere. Right panel: synthesis on the surface of the bodies.}
\label{fig:fig8}
\end{figure}

\begin{figure}
\noindent\includegraphics[width=20pc]{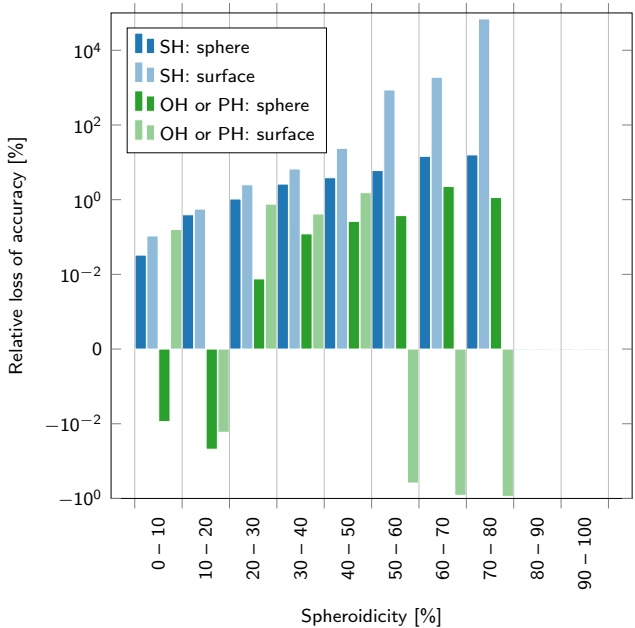}
\caption{Statistical view on the results presented in Fig.~\ref{fig:fig8}. The bars represent mean values of the relative differences in the respective ranges.}
\label{fig:fig9}
\end{figure}

In this work we presented a method to retrieve ellipsoidal harmonics of considerably higher degrees (e.g. $N=500$). Rewriting the computational algorithm in terms of logarithmic expressions eliminates the rather grave limitation of arithmetic overflow. Our tests concerning this matter showed that especially larger objects, say diameters of tens to hundreds of kilometers, are affected by this issue. Following this conclusion, an immediate remedy for this issue can also be achieved in a much simpler manner, i.e. by introducing appropriate units of length in order to reduce the size of the body. Though significant refinement is possible with the scaling approach, the expansion limit is still limited (cf. Fig.~\ref{fig:fig2}). Interestingly enough, the shape seems to influence the maximum computable degree much more for smaller objects. 

Obviously the algorithms to compute the basis functions of EH are very intense in terms of computational complexity. This is even worsened when logarithmic expressions are used. Since many objects are close to the shape of a spheroid, i.e. an ellipsoid of revolution with two equal semi-axes, calculations can be simplified. The oblate and prolate spheroidal harmonics make use of the associated Legendre functions of the first and second kind and are very similar in their handling compared to spherical harmonics. Fast and accurate methods exist to compute the respective basis functions. 

We assessed the suitability of the spherical, spheroidal and ellipsoidal harmonics for modeling the gravitational field of Small Solar System Bodies. On a circumscribed sphere, we conducted closed-loop simulations using polyhedral gravitation formulas to forward-calculate the potential and least-squares algorithms to estimate the respective series coefficients. We reused the estimated coefficients to analyze the effects of divergence by synthesizing and comparing the gravitational potential on the surface of the bodies too. 
In accordance with previous conclusions (e.g. \citet{Garmier2001}, \citet{Hu2015}) our findings imply that the quality of a harmonic approximation depends primarily on the following crucial aspects: the expansion degree $N$, the gravitational signal strength (governed by the distance of the evaluation point from the surface), the geometrical fit of the reference surface and in consequence the distance of the computation points from this surface. Especially the last conclusion might be of practical importance for geophysical inversion applications, e.g. when high-orbit satellite data is included in the gravitational field modeling process. 

The results of the closed-loop simulation studies were extensively demonstrated in the case of Comet 67. This body is in the focus of the public as it is the current target of ESA's space probe Rosetta. The shape of the comet is utterly odd and clearly badly represented by a sphere. Due to its more elongated characteristics also the oblate spheroid is suboptimal. On the other hand, the prolate spheroid turns out to be a very good alternative to the tri-axial ellipsoid. While SH and OH converge rather slowly towards the true potential, the closeness between the decisively more accurate results of PH and EH came as surprise. Especially for the lower degree harmonics, say less than 15, there is virtually no difference between them. For computations on the outside of the Brillouin surfaces, we therefore conclude that PH are to be preferred over EH due to simple mathematics and numerics. A different picture yielded the study of divergence on the comet's surface. In the concave neck region of the body, the spherical and the spheroidal solutions are unable to represent the true potential with acceptable accuracy. The EH handles this topographical depression best, however, it depends on the type of application whether the associated errors of about 10~\% are tolerable or not. For instance, trajectory determination is usually done by means of SH on the outside of the Brillouin sphere and on the basis of the polyhedral shape in close proximity of the body \citep{Scheeres2012}. In order to infer geophysical properties from close-range observations, however, only EH can be trusted.

We repeated the simulation strategy using an extended data set of almost 400 Small Solar System Bodies and found that the majority of the spheroidal solutions (either oblate or prolate, depending on the object's shape) are on average within $\pm 1$~\% of the ellipsoidal's accuracy. Surprisingly, the divergence study resulted in slightly better solutions in OH or PH parameterization for bodies with over 50~\% spheroidicity. The visualization of the relative differences between SH and EH is a striking demonstration of the inappropriateness of the spherical parameterization for highly irregular bodies.

In summary we conclude that spheroidal harmonics should always be considered as an alternative to the much more complicated ellipsoidal parametrization. For instance, the latest findings of NASA's New Horizon mission revealed a very prolate spheroidal shape of Pluto's moon Nix \citep{Stern2015}. Hence, PH might be just the right choice for this object. However, the irregularity of the shapes of these bodies cannot allow for a general statement. In some cases, e.g. for highly elongated bodies or concave geometries, EH might still be the best choice. Using the logarithmic expressions presented in this paper, high resolution fields can be obtained using the ellipsoidal parametrization.

%%% End of body of article:
%
%  ACKNOWLEDGMENTS

\begin{acknowledgments}
We would like to thank the Associate Editor and two anonymous reviewers for critically reading the manuscript and for providing valuable suggestions and corrections to improve this article.
The shape model of Comet 67P/Churyumov-Gerasimenko used in this study is freely available at http://sci.esa.int/jump.cfm?oid=54728.
The data set used in subsection~\ref{subsec:damit} can be obtained as a TAR-GZIP archive from the Database of Asteroid Models from Inversion Techniques (http://astro.troja.mff.cuni.cz/projects/asteroids3D/web.php).
A complete list of numerical results regarding the study of the DAMIT asteroids is available in Table S1 in the supporting information of this article.
\end{acknowledgments}

%% ------------------------------------------------------------------------ %%
%%  REFERENCE LIST AND TEXT CITATIONS
%
% \bibliographystyle{BibTeX/agufull08.bst}
% \bibliography{bib/mybib}

% bbl-file contents
%-------------------------------------------------------------------------------------------------%
% special characters:
% \bibitem[{\textit{Ahipa{\c{s}}ao{\u{g}}lu}(2015)}]{Ahipasaoglu2015}
% Ahipa{\c{s}}ao{\u{g}}lu, S.~D. (2015), {Fast algorithms for the minimum volume
%   estimator}, \textit{Journal of Global Optimization}, \textit{62}(2),
%   351--370, \doi{10.1007/s10898-014-0233-8}.

% \bibitem[{\textit{{\v{D}}urech et~al.}(2010)\textit{{\v{D}}urech, Sidorin, and
%   Kaasalainen}}]{Durech2010}
% {\v{D}}urech, J., V.~Sidorin, and M.~Kaasalainen (2010), {DAMIT: a database of
%   asteroid models}, \textit{Astronomy and Astrophysics}, \textit{513}, A46,
%   \doi{10.1051/0004-6361/200912693}. 

% \bibitem[{\textit{Ritter}(1998)}]{Ritter1998}
% Ritter, S. (1998), {On the Computation of Lam{\'{e}} Functions, of Eigenvalues
%   and Eigenfunctions of Some Potential Operators}, \textit{ZAMM},
%   \textit{78}(1), 66--72,
%   \doi{10.1002/(SICI)1521-4001(199801)78:1\%3C66::AID-ZAMM66\%3E3.0.CO;2-V}.
%-------------------------------------------------------------------------------------------------%

%% ------------------------------------------------------------------------ %%
%
%  END ARTICLE
%
%% ------------------------------------------------------------------------ %%
\end{article}
%
%
%% Enter Figures and Tables here:
%
%
%
\clearpage
%% TABLES
%
\begin{table}
\caption{Exterior solutions to Laplace's equation in four different parametrizations\tablenotemark{a}.}
\centering
\begin{tabular}{c c c c}
\hline
$\left( \xi_1,\xi_2,\xi_3 \right)$  & $\zeta_n$ & $\eta_n$ & $\theta_n$  \\
\hline
$\left( r, \vartheta, \lambda \right)$  & $r^{- \left(n+1\right)}$ & $\overline{P}_{nm} \left( \cos \vartheta \right)$ & $e^{\pm im\lambda}$   \\
$\left( u, \vartheta^{\left(o\right)}, \lambda \right)$  & $Q_{nm} \left( iu/\varepsilon \right)$ & $\overline{P}_{nm} \left( \cos \vartheta^{\left(o\right)} \right)$ & $e^{\pm im\lambda}$   \\
$\left( v, \vartheta^{\left(p\right)}, \lambda \right)$  & $Q_{nm} \left( iv/\varepsilon \right)$ & $\overline{P}_{nm} \left( \cos \vartheta^{\left(p\right)} \right)$ & $e^{\pm im\lambda}$   \\
$\left( \rho,\mu,\nu \right)$  & $F_{nm} \left( \rho \right)$ & $\overline{E}_{nm} \left( \mu \right)$ & $\overline{E}_{nm} \left( \nu \right)$   \\
\hline
\end{tabular}
\tablenotetext{a}{Parametrizations from top to bottom: spherical, oblate spheroidal, prolate spheroidal and ellipsoidal. $\overline{P}_{nm}$ and $Q_{nm}$ are the associated Legendre functions of the first and second kind, respectively. $\overline{E}_{nm}$ and $F_{nm}$ are the two kinds of Lam\'{e} functions. The vinculum indicates full normalization.}
\label{tab:tab1}
\end{table}
\begin{table}
\caption{67P/Churyumov-Gerasimenko characteristics.}
\centering
\begin{tabular}{l l }
\hline
Physical parameters & Value \citep{Sierks2015}\\
\hline
volume & 21.4~km$^3$ \\
density & 470~kg~m$^{-3}$  \\
\hline
Reference surfaces & Dimensions\\
\hline
sphere & 2837~m \\
oblate spheroid & 2862 $\times$ 1920~m \\
prolate spheroid & 2856 $\times$ 2210~m \\
ellipsoid & 2876 $\times$ 2243 $\times$ 1935~m \\
\hline
\end{tabular}
\label{tab:tab2}
\end{table}
\begin{table}
\caption{Statistics of the degree-10 percentage error $\delta_V$ according to SH, OH, PH and EH parametrizations.}
\centering
\begin{tabular}{l c c c }
\hline
& $\min\delta_V$ & $\max\delta_V$ & $\textrm{rms } \delta_V$ \\
\hline
SH & $-3.04 \times 10^{-1}$ & $2.23 \times 10^{-1}$ & $2.64 \times 10^{-2}$ \\
OH & $-2.04 \times 10^{-1}$ & $1.60 \times 10^{-1}$ & $2.07 \times 10^{-2}$ \\
PH & $-1.09 \times 10^{-1}$ & $6.74 \times 10^{-2}$ & $1.12 \times 10^{-2}$ \\
EH & $-8.58 \times 10^{-2}$ & $5.88 \times 10^{-2}$ & $1.09 \times 10^{-2}$ \\
\hline
\end{tabular}
\label{tab:tab3}
\end{table}
\clearpage
% FIGURES
%
\end{document}